\documentclass[iop,twocolappendix]{emulateapj}
\usepackage{apjfonts}
\usepackage{amsmath,amssymb}
\renewcommand{\vec}[1]{\mathbf{#1}}
\newcommand{\diffp}[2]{\frac{\partial #1}{\partial #2}}

\shorttitle{Oblique Modes in the Presence of an Ion Beam}
\shortauthors{D.~Verscharen \& B.~D.~G.~Chandran}

\begin{document}

\title{The Dispersion Relations and Instability Thresholds of Oblique Plasma Modes in the Presence of an Ion Beam}


\author{Daniel Verscharen and Benjamin D.~G.~Chandran \altaffilmark{1}}
\affil{Space Science Center, University of New Hampshire, Durham, NH 03824, USA; daniel.verscharen@unh.edu, benjamin.chandran@unh.edu}
\altaffiltext{1}{Also at Department of Physics, University of New Hampshire, Durham, NH 03824, USA}

\journalinfo{The Astrophysical Journal, 764:88 (12pp), 2013 February 10}
\submitted{Received 2012 November 16; accepted 2012 December 20; published 2013 January 29}

\begin{abstract}
An ion beam can destabilize Alfv\'en/ion-cyclotron waves and
magnetosonic/whistler waves if the beam speed is 
sufficiently large. Numerical solutions of the hot-plasma dispersion
relation have previously shown that the minimum beam speed required to
excite such instabilities is significantly smaller for oblique modes
with $\vec k \times \vec B_0\neq 0$ than for parallel-propagating
modes with $\vec k \times \vec B_0 = 0$, where $\vec k$ is the
wavevector and $\vec B_0$ is the background magnetic field. In this
paper, we explain this difference within the framework of quasilinear
theory, focusing on low-$\beta$ plasmas. We begin by deriving, in the cold-plasma approximation, the
dispersion relation and polarization properties of both oblique and
parallel-propagating waves in the presence of an ion beam. We then
show how the instability thresholds of the different wave branches can
be deduced from the wave--particle resonance condition, the
conservation of particle energy in the wave frame, the sign (positive
or negative) of the wave energy, and the wave polarization. We also
provide a graphical description of the different conditions under
which Landau resonance and cyclotron resonance destabilize
Alfv\'en/ion-cyclotron waves in the presence of an ion beam. We draw
upon our results to discuss the types of instabilities that may limit
the differential flow of alpha particles in the solar wind.
\end{abstract}


\section{Introduction}

The solar wind is a dilute plasma consisting of protons, electrons,
and other ionic species. Among these other ions, the alpha particles
play a special role due to their relatively high abundance. In the fast solar wind, for example, alpha particles comprise $\sim 15-20\%$ of the solar-wind mass
density \citep{bame77}, which makes them dynamically and thermodynamically important.

In situ measurements have shown that the alpha particle velocity
distribution is far from collisionally relaxed.
For example, alpha particles generally have higher temperatures and
outflow velocities than the protons.  Also, like protons, alpha
particles exhibit temperature anisotropy with $T_\perp \neq
T_\parallel$, where $T_{\perp}$ ($T_{\parallel}$) measures the speed
of thermal motions perpendicular (parallel) to the local magnetic
field
\citep{marsch82,vonsteiger95,neugebauer96,reisenfeld01,maruca12}. When
measured in the proton frame, the beam speeds of alpha particles
$U_{\rm i}$ seen in situ at heliocentric distances~$r$ exceeding
0.3~AU is in most cases limited to a value of order the local Alfv\'en
speed $v_{\mathrm A}$. At these heliocentric distances, the values of
both $v_{\rm A}$ and $U_{\rm i}$ decrease with increasing~$r$ while
the proton speed is nearly constant. This means that an ongoing
deceleration mechanism has to act on the alpha particles and regulate
their drift speeds in interplanetary space.

Recent studies~\citep{kasper08,bourouaine11} have shown that~$U_{\rm i}$ is
anti-correlated with the collisional age $r \nu_{\rm col}/U$ of the
solar wind, where $\nu_{\rm col}$ is the Coulomb collision frequency
and $U$ is the proton outflow velocity. Under conditions of rare
collisions in the measured solar wind interval, \cite{bourouaine11}
found an upper threshold of about $1.2v_{\mathrm A}$ for~$U_{\rm i}$
in {\em Helios} data at $r \sim 0.7 \mbox{ AU}$. Several authors \citep[e.g.,][]{isenberg83,gomberoff96b,kasper08,araneda09} have argued that wave--particle interactions would lead to an upper limit of $U_{\mathrm i}\sim v_{\mathrm A}$. The study by \citet{bourouaine11} also found that solar-wind streams
with~$U_{\rm i}$ near this upper limit show excess magnetic fluctuation
power at wavenumbers of up to the inverse local proton
inertial length.

In situ observations suggest that the solar wind hosts a number of
different plasma wave modes, including Alfv\'en waves
\citep{belcher71,tu95}, ion-cyclotron waves \citep{jian10,he12},
whistler waves \citep{podesta11}, and kinetic Alfv\'en waves
\citep{bale05,salem12}.  The properties of these waves depend on the
properties of the plasma in which they propagate. One of these
background properties is the composition of the plasma. Adding other
ion species or relative drifts between the species modifies the waves
significantly
\citep{gomberoff91,gnavi96,gomberoff96,perrone11,marsch11,verscharen11}. The
above-mentioned alpha-particle beam is a typical example of such a
drift. Apart from the modification of dispersion properties, the beam
provides a source of free energy that can lead to the growth of
electromagnetic instabilities, as shown in analytical and numerical
studies \citep[e.g.,][]{montgomery76,gary93,gary00,li00,lu06,lu09,li10}. These drift
instabilities decelerate the drifting particle species to a stable
value, thereby consuming the free energy source.  These treatments
discuss different types of instabilities and determine the threshold
value of the relative drift at which the plasma becomes unstable. Some of the instabilities discussed include a circularly polarized parallel magnetosonic mode, an oblique magnetosonic mode, and two
Alfv\'enic modes.The magnetosonic
mode reaches its maximum growth rate at parallel propagation, at which the 
(previously studied) Alfv\'enic instabilities are suppressed \citep{montgomery76}.
\cite{gary00} have shown that the threshold drift speed for oblique
Alfv\'enic instabilities is significantly smaller than for the parallel
magnetosonic instability --- $U_{\rm i} \gtrsim 0.8 v_{\rm A}$
instead of $U_{\rm i}\gtrsim 1.7 v_{\rm A}$, at least for situations
in which the electron temperature is fairly large (four times the
proton temperature in their study).

The intention of the present article is to give an analytical
explanation for this behavior within the framework of quasilinear
theory. Our primary objective is to elucidate the
physics of these instabilities and not to refine or revise previous
results on the values of these instability thresholds. We base our analysis on the cold-plasma dispersion relation, which we expect provides a reasonably accurate description of the real parts of the wave frequencies when $\beta \ll 1$, where $\beta=8\pi p/B^2$ is the ratio of the plasma pressure $p$ to the magnetic pressure. However, our use of the cold-plasma dispersion relation limits our analysis to the solar corona and the subset of solar-wind streams farther from the Sun that satisfy $\beta \ll 1$. Since most of the solar wind near 1 AU satisfy $\beta\sim 1$, our results do not describe the bulk of near-Earth solar wind. The remainder
of this paper is organized as follows.  In
Section~\ref{sec:properties}, we derive the cold-plasma dispersion
relation for obliquely propagating plasma waves in
the presence of an ion beam. We use the solutions of this dispersion
relation to evaluate the wave polarization and wave energy. In
Sections~\ref{sec:QLT} and~\ref{sec:thresholds} we describe how the
instability thresholds can be determined and understood in terms of
(1) the resonance conditions for wave--particle interactions in
quasilinear theory, (2) the conservation of resonant-particle kinetic
energy in the reference frame moving with the wave along the
background magnetic field, (3) the wave polarization, and (4) the sign
(positive or negative) of the wave energy. In
Section~\ref{sec:discussion} we summarize our findings and discuss
their application to alpha-particle beams in the solar wind.

We note that we do not address the details of the mechanism(s) that
generate ion beams in the solar wind, which could be
cyclotron-resonant wave--particle interactions
\citep[e.g.,][]{mckenzie82,hollweg02,ofman02}, inhomogeneity effects
with low-frequency waves \citep{mckenzie79,isenberg82}, stochastic
heating by low-frequency waves
\citep{mcchesney87,johnson01,chen01,chaston04,chandran10,chandran10b},
and/or some alternative process(es). Also, we concentrate on
instabilities driven by resonant particles and not on the  beam
instabilities that are seen in solutions to the cold-plasma dispersion
relation \citep{gomberoff96}, in part because these ``cold''
instabilities have been found to be stabilized by the presence of
other waves \citep{gomberoff03,araneda04}.

\section{Wave properties in a cold plasma}
\label{sec:properties} 

We consider a uniform, cold plasma consisting of three particle
species: protons, electrons, and a population of beam ions. We assume
that there is a uniform background magnetic field $\vec B_0 =
B_0\hat{\vec e}_z$, and that the electrons and beam ions flow with
respect to the protons at average velocities $U_{\mathrm e}\hat{\vec
  e}_z$ and $U_{\mathrm i}\hat{\vec e}_z$, respectively, with $U_{\mathrm i}>0$ and $B_0>0$.\footnote{In the anti-parallel case ($\vec U_{\mathrm i}\cdot \vec B_0<0$), the same theoretical arguments apply, and the same instabilities are found. Some labels, however, have to be changed in this case.} We carry out
all our calculations in the proton rest frame. We assume that on
average the plasma is charge neutral, with
\begin{equation}
n_{\mathrm p}e+n_{\mathrm i}q_{\mathrm i}-n_{\mathrm e}e=0,
\end{equation}
and carries (on average) zero current,
\begin{equation}
n_{\mathrm i}q_{\mathrm i}U_{\mathrm i}-n_{\mathrm e}eU_{\mathrm e}=0,
\end{equation}
where $n_{\mathrm p}$, $n_{\mathrm i}$, and $n_{\mathrm e}$ are the proton, beam-ion, and electron number densities, and $e$ and $q_{\mathrm i}$ are the proton and beam-ion charges.

\subsection{Dispersion Relation}
We write the Fourier transform of the electric field in terms of its Cartesian components $E_{kx}$, $E_{ky}$, and $E_{kz}$.\footnote{The quantity $\vec E_k$ is the spatial Fourier transform of the electric field, as defined in Equation (\ref{fourierdef}). To obtain the dispersion relation for linear waves, we take $\vec E_k \propto e^{-i\omega t}$.} By linearizing Maxwell's equations and the momentum equations for each  particle species, we obtain an equation of the form
\begin{equation}\label{disptensor}
\frac{\vec kc}{\omega}\times\left( \frac{\vec kc}{\omega}\times \vec E_{k}\right)+\varepsilon \vec E_{k}\equiv \mathcal D\vec E_{k}=0,
\end{equation}
where $\vec k=k_{\perp}\hat{\vec e}_x+k_{\parallel}\hat{\vec e}_z$ is the wavevector, $\omega$ is the frequency,
\begin{equation}\label{full_dieltensor}
\varepsilon=
\begin{pmatrix}
S & -iD & J \\
iD & S & M \\
J& -M & P
\end{pmatrix}
\end{equation}
is the dielectric tensor,
 $S = (R+L)/2$, $D = (R-L)/2$, and the quantities
\begin{equation} 
R =1-\sum\limits_s \left(\frac{\omega_{\mathrm p s}}{\omega}\right)^2\frac{\omega-k_{\parallel}U_{ s}}{\omega-k_{\parallel}U_{s}+\Omega_s}\label{Rdisp} 
\end{equation} 
and
\begin{equation} 
L =1-\sum\limits_s \left(\frac{\omega_{\mathrm p s}}{\omega}\right)^2\frac{\omega-k_{\parallel}U_{ s}}{\omega-k_{\parallel}U_{s}-\Omega_s}
\end{equation} 
extend the notation of \citet{stix92} to account for the relative drift between particle species. The plasma frequency of species $s$ is given by $\omega_{\mathrm ps}^2\equiv 4\pi n_sq_s^2/m_s$. The indices $\perp$ and $\parallel$ refer to the directions perpendicular and parallel to the background magnetic field. The gyrofrequency of the corresponding particle species is denoted $\Omega_s\equiv q_sB/(m_sc)$ and is negative for electrons. The terms
\begin{equation} 
J =-\sum\limits_s \left(\frac{\omega_{\mathrm p s}}{\omega}\right)^2k_{\perp}U_s \frac{\omega-k_{\parallel}U_s}{\left(\omega-k_{\parallel}U_s\right)^2-\Omega_s^2}
\end{equation} 
and
\begin{equation} 
M =i\sum\limits_s \left(\frac{\omega_{\mathrm p s}}{\omega}\right)^2k_{\perp}U_s \frac{\Omega_s}{\left(\omega-k_{\parallel}U_s\right)^2-\Omega_s^2}\label{Mdisp}
\end{equation} 
represent new components of the dielectric tensor that are absent when the beam is absent, and
\begin{equation}
P=1-\sum\limits_{s}\left(\frac{\omega_{\mathrm p s}}{\omega}\right)^2\left[\frac{\omega^2}{\left(\omega-k_{\parallel}U_s\right)^2}+\frac{k_{\perp}^2U_s^2}{\left(\omega-k_{\parallel}U_s\right)^2-\Omega_s^2}\right].
\end{equation}
The term ``1'' in the expressions for $R$, $L$, and $P$ stems from the displacement current and can be neglected in our study.

To obtain the dispersion relation, we set
\begin{equation}
\det \mathcal D = 0.
\label{eq:DR} 
\end{equation} 
We solve Equation~(\ref{eq:DR}) in two steps. First, we obtain
approximate solutions to the dispersion relation in the limit that
$m_{\mathrm e} \rightarrow 0$, where $m_{\rm e}$ is the electron
mass. Second, we use this approximate solution as the starting point
for a Newton's-method solution to the full cold-plasma dispersion
relation, taking into account the finite value of $m_{\rm e}/m_{\rm p}$.

In the limit $m_{\mathrm e} \rightarrow 0$, Equations (\ref{Rdisp}) through (\ref{Mdisp}) become
\begin{align}
R&\simeq \frac{\omega_{\mathrm{pp}}^2}{\Omega_{\mathrm p}\left(\Omega_{\mathrm p}+\omega\right)}+\frac{\omega_{\mathrm{pi}}^2\left(\omega-k_{\parallel}U_{\mathrm i}\right)^2}{\omega^2\Omega_{\mathrm i}\left(\Omega_{\mathrm i}+\omega-k_{\parallel}U_{\mathrm i}\right)},\\
L&\simeq \frac{\omega_{\mathrm{pp}}^2}{\Omega_{\mathrm p}\left(\Omega_{\mathrm p}-\omega\right)}+\frac{\omega_{\mathrm{pi}}^2\left(\omega-k_{\parallel}U_{\mathrm i}\right)^2}{\omega^2\Omega_{\mathrm i}\left(\Omega_{\mathrm i}-\omega+k_{\parallel}U_{\mathrm i}\right)},\\
J&\simeq -\left(\frac{\omega_{\mathrm{pi}}}{\omega}\right)^2 k_{\perp}U_{\mathrm i}\frac{\omega-k_{\parallel}U_{\mathrm i}}{\left(\omega-k_{\parallel}U_{\mathrm i}\right)^2-\Omega_{\mathrm i}^2},
\end{align}
and
\begin{equation}
M\simeq i \left(\frac{\omega_{\mathrm {pi}}}{\omega}\right)^2\frac{k_{\perp}U_{\mathrm i}}{\Omega_{\mathrm i}}\frac{\left(\omega-k_{\parallel}U_{\mathrm i}\right)^2}{\left(\omega-k_{\parallel}U_{\mathrm i}\right)^2-\Omega_{\mathrm i}^2}.
\end{equation}
 As $m_{\mathrm e}\rightarrow 0$, $P\rightarrow
\infty$, and only those terms in $\det \mathcal D$ that contain a
factor of~$P$ need to be retained when solving
Equation~(\ref{eq:DR}). After factoring out this factor of~$P$ from
Equation~(\ref{eq:DR}), we obtain the ($m_e\rightarrow 0$) dispersion
relation for Alfv\'en/ion-cyclotron and fast/whistler waves,
\begin{equation}\label{fulldisp}
\left(\frac{kc}{\omega}\right)^4\cos^2\theta -\left(\frac{kc}{\omega}\right)^2\left(\frac{R+L}{2}\right)\left(1+\cos^2\theta \right)+RL=0,
\end{equation}
where $\theta$ is the angle between $\vec k$ and $\vec B_0$. When $\theta=0$, Equation~(\ref{fulldisp}) becomes
\begin{equation}
\left[\left(\frac{kc}{\omega}\right)^2-L\right]\left[\left(\frac{kc}{\omega}\right)^2-R\right]=0.
\label{eq:parallel}
\end{equation}
Each of the equations $(kc/\omega)^2-L=0$ and $(kc/\omega)^2 - R= 0$ 
leads to a third-order polynomial equation for $\omega$. When the beam
ions are alpha particles, the equation $(kc/\omega)^2-L=0$ is
equivalent to Equation~(2) of \citet{gomberoff91}.  When $\theta\neq 0$,
Equation~(\ref{fulldisp}) leads to a sixth-order polynomial equation for
$\omega$, which we solve numerically using the Laguerre method
\citep{press92}.

As mentioned above, we use our solution of Equation~(\ref{fulldisp})
as the starting point for a Newton's-method solution of
Equation~(\ref{eq:DR}) taking into account the finite electron
mass. This second step to our solution of Equation~(\ref{eq:DR}) leads
to only minor refinements to the wave frequency. However, it enables
us to calculate polarization variables for each mode such as $E_{k
  z}/E_{kx}$, which are relevant to our discussion of Landau damping
in the sections to follow.

For the remainder of this paper, we use the full solution of the
cold-plasma dispersion relation with finite~$m_{\rm e}$ rather than
working in the zero-$m_{\rm e}$ limit. We also from here on restrict
ourselves to the case in which the beam ions are alpha particles. We
define
\begin{equation}
\eta\equiv \frac{n_{\mathrm i}}{n_{\mathrm p}}
\end{equation}
and
\begin{equation}
v_{\mathrm A} \equiv \frac{B_0}{\sqrt{4\pi n_{\mathrm p}m_{\mathrm p}}}.
\end{equation}
\begin{figure*}
\epsscale{1.}
\includegraphics[width=\textwidth]{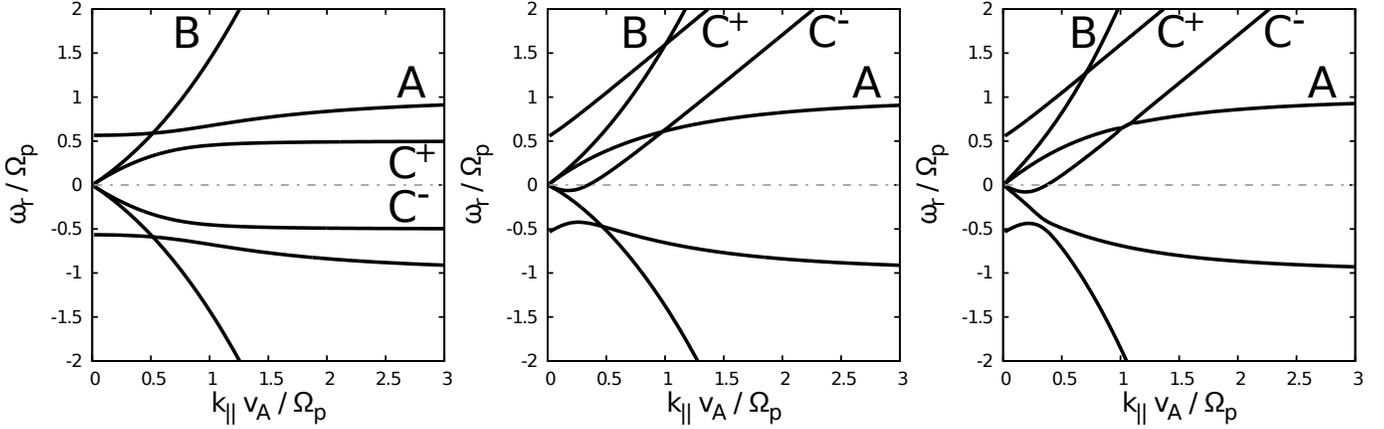}
\caption{Solutions to Equation~(\ref{disptensor}) for an alpha-particle beam. The letters are explained in the text. For the left panel, $\eta=0.075$, $\theta=0^{\circ}$, and $U_{\mathrm i}=0$. For the middle panel, $\eta=0.05$, $\theta=1.5^{\circ}$, and $U_{\mathrm i}=1.1v_{\mathrm A}$. For the right panel, $\eta=0.05$, $\theta=45^{\circ}$, and $U_{\mathrm i}=1.1v_{\mathrm A}$. \label{fig_dispersion}}
\end{figure*} 

In Figure~\ref{fig_dispersion}, we plot
$\mathrm{Re}\,\omega=\omega_{\mathrm r}$ for each of the six roots to
Equation~(\ref{fulldisp}) for three different choices of $\eta$,
$U_{\mathrm i}$, and $\theta$. We focus on four of these branches,
which we label according to their asymptotic behaviors at
large~$k_\parallel$. Branch A is the proton-cyclotron branch, for
which $\omega \rightarrow \Omega_{\rm p}$ as $k_\parallel \rightarrow
\infty$. Branch~B is the fast-magnetosonic/whistler branch, which
behaves like a fast magnetosonic wave at $k v_{\rm A}/\Omega_{\rm p} <
1$ and a whistler wave at $k v_{\rm A} /\Omega_{\rm p} >
1$. Branch~$\mbox{C}^+$ is the forward-propagating alpha-cyclotron
branch, for which $\omega \rightarrow k_\parallel U_{\rm i} +
\Omega_{\rm i}$ at large~$k_\parallel$. In the frame of the alpha
particles, the Doppler-shifted frequency of this wave $\omega -
k_\parallel U_{\rm i}$ approaches~$\Omega_{\rm i}$ as $k_\parallel
\rightarrow \infty$. Branch~$\mbox{C}^-$ is what we call the
``counter-propagating'' alpha-cyclotron branch.  In the alpha-particle
frame, this wave propagates in the~$-z$ direction and has a
Doppler-shifted frequency $\omega - k_\parallel U_{\rm i} $ that
approaches~$- \Omega_{\rm i}$ as $k_\parallel \rightarrow \infty$.  We
note that for each case shown, $\mathrm{Im}\,\omega=\gamma=0$ except
at the intersections of different branches of the dispersion relation, where two branches have equal $\omega_{\mathrm r}$. In the right panel of Figure~\ref{fig_dispersion}, this occurs, e.g., at $k_{\parallel} v_{\mathrm A}/\Omega_{\mathrm
  p}\approx 1.1 $ and $\omega_{\mathrm r} \approx 0.7 \Omega_{\mathrm p}$, where
$\gamma\simeq \pm 0.02\Omega_{\mathrm p}$.  Both solutions have the same absolute value of $\gamma$ but with opposite signs. Other modes can also merge in certain wavenumber ranges for different parameter choices. The magnetosonic mode B and the counter-propagating alpha-cyclotron mode C$^-$ merge at higher drift speeds and show also a non-zero growth rate in the wavenumber range where they have equal $\omega_{\mathrm r}$. This unstable behavior of the cold-plasma dispersion relation is known
as the ion--ion resonant instability \citep{gnavi96,gomberoff96}.

\subsection{Wave Polarization}

\begin{figure}
\epsscale{1.}
\includegraphics[width=\columnwidth]{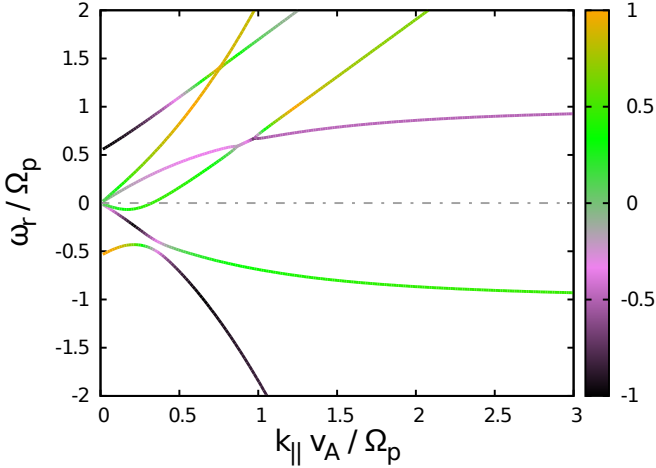}
\caption{The dispersion relation is shown for $\theta=45^{\circ}$, $\eta=0.05$, and $U_{\mathrm i}=1.2v_{\mathrm A}
$. The color coding represents the polarization function $\mathcal P$. All modes in this diagram are elliptically polarized at oblique propagation. \label{fig_polarization}}
\end{figure}

The sense of circular or elliptical polarization of the waves can be defined in terms of the quantity
\begin{equation}
\mathcal P \equiv \frac{|E_{k,\mathrm r}|-|E_{k,\mathrm l}|}{|E_{k,\mathrm r}|+|E_{k,\mathrm l}|}=\frac{|\mathcal A-i|-|\mathcal A+i|}{|\mathcal A-i|+|\mathcal A+i|},
\label{eq:defP} 
\end{equation}
where $E_{k,\mathrm r}\equiv(E_{kx}-iE_{ky})/\sqrt{2}$ and
$E_{k,\mathrm l}\equiv(E_{kx}+iE_{ky})/\sqrt{2}$.  With the use of the $x$ and
$y$ components of Equation~(\ref{disptensor}), we can write
\begin{equation}
\mathcal A \equiv \frac{(S-n_x^2-n_z^2)(J+n_xn_z)+iDM}{(S-n_z^2)M-iD(J+n_xn_z)},
\end{equation}
where $n_x = (kc/\omega) \sin \theta$ and $n_z =
(kc/\omega) \cos\theta$.
When $\theta = 0$, we can use just the $x$ component of Equation~(\ref{disptensor}) to obtain the simpler expression
\begin{equation}
\mathcal A \equiv \frac{iD}{S-n_z^2}.
\label{eq:Apar} 
\end{equation}
Following the standard convention, we say that a wave is left (right)
circularly polarized if the electric field vector in the $x$-$y$ plane
rotates in the same sense as the cyclotron motion of a proton
(electron).  The connection between left/right circular polarization
and the value of $\mathcal P$ depends upon the sign of the real part
of the wave frequency~$\omega_{\rm r}$.  When $\omega_{\rm r} >0$,
left-hand circular polarization corresponds to $\mathcal P = -1$,
whereas right-hand circular polarization corresponds to $\mathcal P =
1$. When $\omega_{\rm r} < 0$, it is the reverse: left-hand circular
polarization corresponds to $\mathcal P = 1$ and right-hand circular
polarization corresponds to $\mathcal P = - 1$. In general, waves are
not circularly polarized, but elliptically polarized with $\mathcal P
\neq \pm 1$. Linear polarization corresponds to~$\mathcal
P= 0$. The value of $\mathcal P$ is shown in Figure~\ref{fig_polarization} for the case of oblique propagation.

It is instructive to consider the sense of polarization of the wave
branches A, B, and $\mbox{C}^\pm$ in Figure~\ref{fig_dispersion} when
$\theta = 0$. It can be seen from Equations~(\ref{eq:parallel}) and
(\ref{eq:Apar}) that $\mathcal P = \pm 1$ when $\theta = 0$, so that
these wave branches are circularly polarized.  More specifically,
branch~B is right-circularly polarized, and branches A and $\mbox{C}^+$ are left-circularly polarized. Branch $\mbox{C}^-$ is also left circularly
polarized when $U_{\rm i} = 0$, or, if $U_{\rm i} > 0$, when
$\omega_{\rm r} < 0$. However, when $U_{\rm i} > 0$ and $\omega_{\rm
  r} > 0$, branch $\mbox{C}^-$ is right-circularly polarized. It is
important to recall, however, that we have defined the sense of
polarization in the rest frame of the protons. If we were to switch
into the rest frame of the alpha particles, the Doppler-shifted wave
frequency $\omega_{\rm r} - k_\parallel U_{\rm i}$ of branch
$\mbox{C}^-$ would always be negative, and in this reference frame
branch $\mbox{C}^-$ would appear to be left circularly polarized.

The longitudinal polarization is important for Landau-resonant interactions.
We define the longitudinal polarization as 
\begin{equation}\label{mathcalpz}
\mathcal P_z\equiv \left[\left|\frac{E_{kx}}{E_{kz}}\right|+\left|\frac{E_{ky}}{E_{kz}}\right|\right]^{-1}.
\end{equation}
The right-hand side of Equation~(\ref{mathcalpz}) can be evaluated with the
use of the first two lines of Equation~(\ref{disptensor}).  It is found to
be always much less than one for the treated modes. Its maximum value
in the treated frequency and wavenumber range is of the order of
$\mathcal P_z\lesssim 10^{-3}$. In parallel propagation, the
determination of the parallel component of the electric field is not possible anymore via
the first two components of Equation~(\ref{disptensor}) since the
dielectric tensor becomes block diagonal. The only parallel mode with
$E_{kz}\neq 0$ is the $P=0$ mode, which has very high frequencies for
the plasma conditions that we consider, and which is excluded
here. Therefore, we can assume all parallel modes to have $E_{kz}=0$.

\subsection{Wave Energy}
\label{sec:W} 

The wave energy density per unit volume in $k$ space, denoted $W$, is given by
\begin{equation}\label{waveenerg}
\left(2\pi\right)^3 W=\frac{1}{8\pi}\left.\left[\vec B^{\ast}_{k}\cdot\vec B_{k}+\vec E^{\ast}_{k}\cdot \diffp{}{\omega}(\omega \varepsilon_{\mathrm h})\vec E_{k}\right]\right|_{\omega=\omega_{\mathrm r}},
\end{equation}
where $\vec E_{k}$ ($\vec B_{k}$) is the Fourier transform of the fluctuating electric (magnetic) field, and $\varepsilon_{\mathrm h}=(\varepsilon+\varepsilon^{\dagger})/2$ is the hermitian part of the dielectric tensor \citep[][see also the discussion following Equation 27 of \citealt{chandran10b}]{stix92}.  The dagger symbol denotes the adjoint matrix. The factor of $\left(2\pi \right)^3$ on the left-hand side of Equation~(\ref{waveenerg}) does not appear in Equation~(20) of Chapter 4 of \citet{stix92}. This difference arises because we define
\begin{equation}\label{fourierdef}
\vec E_{k}=\int \vec E(\vec x, t)e^{-i\vec k\cdot \vec x}\mathrm d^3 x,
\end{equation}
whereas \citet{stix92} defines
\begin{equation}
\vec E_{k}=\left(2\pi \right)^{-3/2} \int \vec E(\vec x, t)e^{-i\vec k\cdot \vec x}\mathrm d^3 x.
\end{equation}

This energy density describes both the energy density of the electromagnetic wave fields and the kinetic energy density of the particles under the assumption of a slow change of the wave amplitude with time.

\begin{figure}
\epsscale{1.}
\includegraphics[width=\columnwidth]{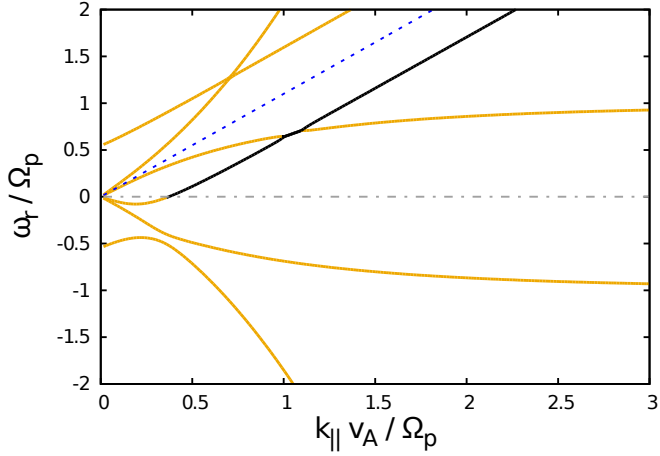}
\caption{Orange (black) color curves indicate positive (negative) values of the wave energy $W$ for the different solutions to the dispersion relation for  $\theta=45^{\circ}$, $\eta=0.05$, and $U_{\mathrm i}=1.1v_{\mathrm A}$. The blue-dotted line indicates $\omega_{\mathrm r}=k_{\parallel}U_{\mathrm i}$. \label{fig_negenerg}}
\end{figure}

It was found by \citet{kadomtsev65} that the term containing the
derivative of the dielectric tensor in the expression for the wave
energy can be negative and even lead to a situation in which waves
have \emph{negative energy}, meaning that $W<0$.  In this case, the
averaged kinetic energy of the particles plus the energy of the
electromagnetic field is larger when the wave is absent than when it
is present~\citep{dawson61,hollweg71}. The behavior of negative-energy
waves under the influence of dissipation is
counter-intuitive. Dissipation, which removes energy from the
fluctuations and converts it into heat, causes negative-energy waves
to grow \citep{hall66,hollweg71}.  Similarly, if a negative-energy
wave loses energy through a parametric instability, its own amplitude
grows, increasing the rate of this instability. This is the so-called
\emph{explosive instability} \citep{oraevsky84}. We find that the
counter-propagating alpha-cyclotron wave C$^-$ has negative energy for all
$\omega_{\rm r}>0$. This situation is depicted in
Figure~\ref{fig_negenerg}. Negative-energy waves in the presence of an
ion beam have been investigated previously by \citet{gnavi96}.

In all cases that we have investigated so far, when the
alpha-cyclotron branch merges with the proton-cyclotron or the
magnetosonic mode, the resulting branches have negative energy
throughout the finite interval of~$k_\parallel$ in which the two
branches have the same value of~$\omega_{ \rm r}$. This situation can
be seen in Figure~\ref{fig_negenerg}, where the forward propagating
proton-cyclotron wave (branch A) and the alpha-cyclotron wave (branch C$^-$) that propagates
backwards in the frame of the alpha particles merge.

\section{Quasilinear Theory}
\label{sec:QLT} 

Quasilinear theory provides an approximate, analytical description of wave--particle interactions in a collisionless plasma. One of the central assumptions of this theory is that the fluctuation amplitudes are sufficiently small  that they can be described as a superposition of linear waves, and a particle's orbit can be approximated as an unperturbed helix aligned with the background magnetic field.  In addition, the theory assumes that the growth or damping rates of the waves are much smaller than the linear wave frequencies. 
The starting point of this theory is the Vlasov equation,
\begin{equation}
\diffp{f_s}{t}+\vec v\cdot \diffp{f_s}{\vec x}+\frac{q_s}{m_s}\left(\vec E+\frac{1}{c}\vec v\times \vec B\right)\cdot \diffp{f_s}{\vec v}=0,
\end{equation}
which describes the evolution of the distribution function $f_s(\vec x,\vec v,t)$ of particle species $s$ with mass $m_s$ and charge $q_s$ under the influence of the electric field~$\vec E$ and magnetic field~$\vec B$.
In a constant homogeneous background magnetic field ($\vec B_0=B_0\hat{\vec e}_z$), a perturbation analysis leads to the following relation for the slow temporal evolution of the distribution function:
\begin{multline}
\diffp{f_s}{t}=\lim _{V\to \infty}\sum \limits_{n=-\infty}^{+\infty}\frac{q_s^2}{8 \pi^2m_s^2}\int \frac{1}{Vv_{\perp}}Gv_{\perp}\delta(\omega_{k\mathrm r}-k_{\parallel}v_{\parallel}-n\Omega_s)\times \\
\times \left|\psi_{n,k} \right|^2Gf_s\mathrm d^3k,\label{qldiff}
\end{multline}
where
\begin{equation}
G\equiv\left(1-\frac{k_{\parallel}v_{\parallel}}{\omega_{k\mathrm r}}\right)\diffp{}{v_{\perp}}+\frac{k_{\parallel}v_{\perp}}{\omega_{k\mathrm r}}\diffp{}{v_{\parallel}}
\end{equation}
and 
\begin{multline}
\psi_{n,k}\equiv \frac{1}{\sqrt{2}}\left[E_{k,\mathrm r}e^{i\phi}J_{n+1}(\sigma_s)+E_{k,\mathrm l}e^{-i\phi}J_{n-1}(\sigma_s)\right]\\
+\frac{v_{\parallel}}{v_{\perp}}E_{kz}J_n(\sigma_s)
\label{eq:psink} 
\end{multline}
\citep{kennel66,stix92}.  The azimuthal angle of the
wavevector $\vec k$ is defined as $\phi$. The parallel and perpendicular components of the velocity in cylindrical coordinates are denoted $v_{\parallel}$ and $v_{\perp}$, respectively. The real part of the
frequency that is a solution to the dispersion relation is denoted 
$\omega_{k \mathrm r}$. The amplitude function $\psi_{n,k}$ contains
the Bessel function of $n$th order $J_n$ with the  argument
$\sigma_s\equiv k_{\perp}v_{\perp}/\Omega_s$. The quantity~$V$ is a
volume that is related to our Fourier transform convention. We treat
the plasma as infinite and homogeneous. In
order to make the Fourier transform of a function $h(\vec x)$
converge, we first multiply this function by a window function $W$
that is $1$ inside a cubical region of volume~$V$ centered on the
origin and $0$ outside this volume. We then take the limit of~$V\rightarrow \infty$.

Equation~(\ref{qldiff}) shows that, in the framework of quasilinear theory, waves cause particles to diffuse in velocity space. Because of the delta function in Equation~(\ref{qldiff}), particles of species~$s$ with velocity $\vec v$ are affected by waves with wavevector $\vec k$ and frequency $\omega_{k\mathrm r}$ only if the waves and particles satisfy the resonance condition
\begin{equation}\label{resecond}
\omega_{k\mathrm r}-k_{\parallel}v_{\parallel}=n\Omega_s,
\end{equation}
where $n$ is any integer. Equation~(\ref{resecond}) is called the Landau-resonance condition when $n=0$ and the cyclotron-resonance condition when $n\neq 0$. If there are only waves at a single $\vec k$ and $\omega_{k\mathrm r}$, then the diffusive flux of particles in velocity space is nonzero only within narrow bands of $v_{\parallel}$ that correspond to the solutions of Equation~(\ref{resecond}). Within those bands, the diffusive flux of particles is tangent to semicircles in the $v_{\parallel}$-$v_{\perp}$ plane defined by the equation
\begin{equation}
\left(v_{\parallel}-v_{\mathrm{ph}}\right)^2+v_{\perp}^2=\mathrm{constant},
\label{eq:constE} 
\end{equation}
where $v_{\mathrm{ph}}\equiv\omega_{k\mathrm r}/k_{\parallel}$ is the parallel phase velocity of the waves \citep{kennel66}. When a spectrum of waves is present, velocity-space diffusion is not necessarily limited to narrow bands of $v_{\parallel}$ values, and particles can instead diffuse throughout an extended region in velocity space. For example, energetic protons undergoing cyclotron interactions with a broad spectrum of non-dispersive ($k_{\parallel}\ll \Omega_{\mathrm p}/v_{\mathrm A}$) Alfv\'en waves with dispersion relation $\omega_{k\mathrm r}=k_{\parallel}v_{\mathrm A}$ scatter along extended semi-circular arcs centered on $v_{\mathrm A}\hat{\vec e}_z$. In contrast, when wave--particle interactions are limited to the Landau resonance, particles can diffuse over an extended range in velocity space only if the spectrum of waves has a continuous range of $v_{\mathrm{ph}}$ values, because the Landau resonance only arises when $v_{\parallel}=v_{\mathrm{ph}}$. At each resonant value of $v_{\parallel}$, Landau-resonant interactions lead to diffusion only in $v_{\parallel}$, not in $v_{\perp}$, and so the extended diffusion paths for Landau-resonant interactions with a spectrum of waves are horizontal lines in the $v_{\parallel}$-$v_{\perp}$ plane. For our purposes, it is sufficient to consider the localized velocity-space diffusion that results from waves at a single $\vec k$ and $\omega_{k\mathrm r}$. In this case, Equation~(\ref{eq:constE}) determines the vector direction of the diffusive flux, up to a 180 degree ambiguity. This ambiguity can be resolved by noting that the net diffusive particle flux is directed from regions of higher particle concentration to regions of lower particle concentration in velocity space.

 Bessel functions have
the property $J_n(0)=\delta_{n,0}$, which shows that, for ``slab'' waves with
$k_{\perp}=0$, resonances occur only when $n=-1$, 0, or 1 . We assume that 
\begin{equation}
\sigma_s \ll 1,
\label{eq:sigma} 
\end{equation} 
set
\begin{equation}
J_n(\sigma_s) \rightarrow \delta_{n,0},
\label{eq:Jn0} 
\end{equation} 
and hence ignore  resonances with $|n| > 1$. Equation~(\ref{eq:sigma})  is satisfied for Alfv\'en/ion-cyclotron waves and magnetosonic/whistler waves provided $\omega_{k\mathrm r} \lesssim \Omega_{\rm p}$, 
$\beta \ll 1$, and $\theta$ is not too close to~$90^\circ$.

\begin{figure}
\epsscale{1.}
\includegraphics[width=\columnwidth]{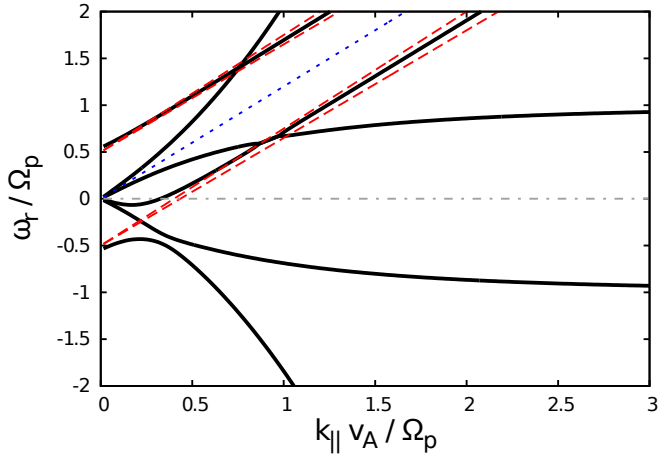}
\caption{Dispersion relation (black, solid) and cyclotron-resonance condition (red, dashed) for  $\theta=45^{\circ}$, $U_{\mathrm i}=1.2v_{\mathrm A}$, and $\eta=0.05$. The blue dotted line represents $\omega_{\mathrm r}=k_{\parallel}U_{\mathrm i}$. The triple-dotted gray line corresponds to $\omega_{\mathrm r}=0$. The spread of the resonance lines corresponds to a parallel ion beta of $\beta_{\parallel\mathrm i}\simeq 10^{-4}$ for this choice of $\eta$.\label{fig_alpha_res}}
\end{figure}

Throughout this paper, we work in the proton rest frame. This means
that the parallel particle velocity $v_{\parallel}$ in the resonance
condition (\ref{resecond}) for the beam particles has two
contributions, namely the beam velocity and the offset from the beam
center. In some of the figures, we will plot ``resonance lines'',
which represent the condition $\omega_{\rm r} - k_\parallel
v_\parallel = n \Omega_{\mathrm i}$. A solution of the resonance condition
(Equation~(\ref{resecond})) corresponds to an intersection of a
resonance line and a plot of a dispersion relation $\omega_{\rm r} =
\omega_{k\mathrm r}$. Since the distribution function of the alpha particles
has a finite thermal width, we also plot a spread of the resonance
lines taking particles into account that are slightly faster or slower
than the bulk speed of the beam. The two resonance lines with the same $n$ then only represent the approximate upper and lower limits of parallel particle speeds that are available in the distribution function. Generally, there are always particles that can fulfill the resonance condition in the area between both plotted resonance lines with equal $n$. This is the only point in our
considerations where thermal effects play a role, since we employ the
cold-plasma dispersion relation. Some potential resonances are shown
in Figure~\ref{fig_alpha_res}.
Here we plot the six solutions in this wavenumber/frequency range  to the dispersion relation,
Equation (\ref{disptensor}). The red lines show the first two ($n=\pm 1$)
cyclotron resonances for alpha particles with parallel velocities
$v_{\parallel}=U_{\mathrm i}\pm 0.05v_{\mathrm A}$. We take
$\Delta v_{\parallel}=\left|v_{\parallel}-U_{\mathrm i}\right|$ to correspond to twice the parallel thermal speed $v_{\mathrm{thi}}\equiv\sqrt{2k_{\mathrm B}T_{\parallel \mathrm i}/m_{\mathrm i}}$ of the alpha particles on the assumption that there are too few alpha particles with $\Delta v_{\parallel}>2v_{\mathrm{thi}}$ to drive a significant instability. This situation corresponds to a plasma with 
\begin{equation}
\beta_{\parallel \mathrm i}\equiv \frac{8\pi n_{\rm i} k_{\rm B}
  T_{\parallel \rm i}}{B_0^2} = \left(\frac{n_{\rm i}m_{\rm
    i}}{n_{\rm p}m_{\rm p}}\right) \left(\frac{v_{ \mathrm {thi}}}{v_{\rm A}}\right)^2
\simeq  10^{-4}
\end{equation}
where $T_{\parallel \rm i}$ is the
parallel ion (alpha-particle) temperature.  The Landau resonance,
which is only acting if $E_{kz}\neq 0$ or $k_\perp \neq 0$, is located on resonance lines with a
small spread around the line $\omega_{\mathrm
  r}=k_{\parallel}U_{\mathrm i}$, which corresponds to the vicinity of
the blue-dotted line in Figure~\ref{fig_alpha_res}.

\section{Instability Thresholds}\label{sect_inst}
\label{sec:thresholds} 

\begin{figure}
\epsscale{.80}
\plotone{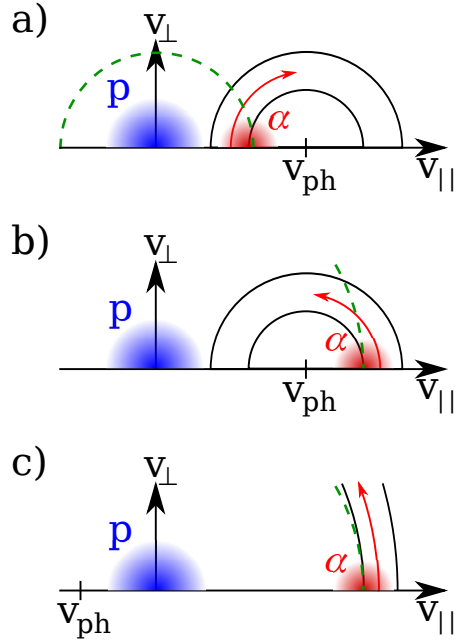}
\caption{Direction of the diffusive particle flux for cyclotron-resonant alpha particles. The regions occupied by thermal protons (alpha particles) are shown as blue (red) filled circles. The arrows and circles show the diffusion paths around the position of the parallel phase speed $v_{\mathrm{ph}}$ for a cyclotron resonance. The green dashed lines are iso-contours of the kinetic energy in the proton frame. a) $U_{\mathrm i}<v_{\mathrm{ph}}$, b) $U_{\mathrm i}>v_{\mathrm{ph}}$, c) $v_{\mathrm{ph}}<0$.  \label{fig_diff_paths}}
\end{figure}

\begin{figure}
\epsscale{.80}
\plotone{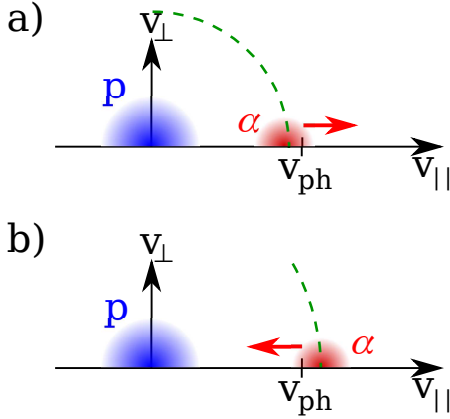}
\caption{Direction of the diffusive particle flux for Landau-resonant alpha particles. The regions occupied by thermal protons (alpha particles) are shown as blue (red) filled circles. The arrows show the diffusion paths at the position of the parallel phase speed $v_{\mathrm{ph}}$ for a Landau resonance. The green dashed lines are iso-contours of the kinetic energy in the proton frame. a) $U_{\mathrm i}<v_{\mathrm{ph}}$, b) $U_{\mathrm i}>v_{\mathrm{ph}}$.  \label{fig_diff_paths_landau}}
\end{figure}

In this paper, we focus on instabilities driven by the velocity-space
diffusion of alpha particles that interact resonantly with the
unstable wave.   Some possible velocity-space diffusion paths for a beam of
alpha particles are shown in Figure~\ref{fig_diff_paths} for the cyclotron-resonant case and in Figure~\ref{fig_diff_paths_landau} for the Landau-resonant case based on the assumption that the particle distribution functions are isotropic about the mean velocity of the particle species.
The resonant alpha particles diffuse down the density gradient in
velocity space, from larger particle concentrations towards smaller
particle concentrations along the diffusion paths.  We concentrate on
the normal case with positive wave energy first.  In this case, if the
particles gain energy during the resonant quasilinear diffusion, the
energy is taken from the wave, and the wave is damped. On the other
hand, if the particles diffuse towards smaller particle energies, then
the energy lost by the particles is given to the wave, leading to wave
amplification.  To see whether particles gain or lose energy, the
 quasilinear-diffusion paths can be compared with curves of
constant particle energy, which are circles centered on the origin in
the $v_{\parallel}$-$v_{\perp}$ plane. 
 The key point demonstrated by
Figure~\ref{fig_diff_paths} is that cyclotron-resonant wave--particle
interactions cause alpha particles to lose energy if and only if $0 <
v_{\mathrm{ph}} < U_{\mathrm i}$. Therefore, if $W>0$,
the wavevectors and frequencies of unstable waves must lie above the
line $\omega_{\rm r} = 0$ and below the line $\omega_{\mathrm
  r}=k_{\parallel}U_{\mathrm i}$ in the $k_\parallel$-$\omega_{\mathrm r}$ plane. Figure~\ref{fig_diff_paths_landau} demonstrates that the same requirement is also valid for Landau-resonant particles.  A mathematical proof of the instability condition $0<v_{\mathrm{ph}}<U_{\mathrm i}$ is given in the Appendix.

For simplicity, we take the protons to be Maxwellian for the purposes
of inferring the effects of wave--particle interactions on the
waves. We thus neglect proton temperature anisotropy and proton
beams. As a consequence, if protons diffuse in velocity space from
regions of large particle concentration towards regions of small
particle concentration, then they must gain energy. Because the
protons are the majority ion species, we expect that proton damping will dominate
over the alpha-particle instability drive if thermal protons can
resonate with a wave.

The foregoing discussion and the condition $\sigma_{s} \ll 1$
imply that a solution of the dispersion relation with $\vec k=\vec k_1$, $\omega_{k\mathrm r}=\omega_1$, and $W>0$  is driven unstable by resonant interactions with
an alpha-particle beam that is isotropic about the beam velocity $U_{\mathrm i}$ in $v$-space if and only if the following conditions are
satisfied:
\begin{enumerate}
\item The coordinates ($k_{1\parallel}$, $\omega_1$) lie between two resonance lines with equal $n$ with $n=+1$, $n=0$, or $n=-1$. If ($k_{1\parallel}$, $\omega_1$) lies between two resonance lines with $n=-1$, then we say that the instability is driven by an $n=-1$ resonance, and likewise for $n=0$ and $n=+1$.
\item For an instability driven by an $n=+1$ resonance, 
  $E_{k,\rm l}$ must be nonzero (i.e.,
 $\mathcal P\neq +1$). For an instability driven by an $n=-1$ resonance,
$E_{k, \rm r}$ must be nonzero (i.e., $\mathcal P\neq
  -1$).  For an instability driven by an $n=0$ resonance, $E_{kz}$ or $k_\perp$ must be nonzero.
\item The parallel phase velocity of the waves is between zero and the
  beam speed: $0<\omega_{1}/k_{1\parallel}<U_{\mathrm i}$.
\item The wave at $\vec k_{1}$ and $\omega_1$ does not satisfy the resonance condition Equation~(\ref{resecond})  with thermal protons.
\end{enumerate}
Condition~2 above on the wave polarization follows from
substituting Equation~(\ref{eq:Jn0}) into Equation~(\ref{eq:psink}).
Importantly, a mode with $W<0$ that satisfies the above criteria is
damped, because an increase in the energy of a negative-energy wave
corresponds to a decrease in the wave's amplitude.
In the next two subsections, we use the above criteria to determine which branches of the
dispersion relation are unstable and at which wavenumbers as a
function of the beam velocity, the angle of propagation, and the
parallel thermal widths of the alpha and proton distributions.

\subsection{Parallel Case~($k_\perp = 0$)}
\label{sec:parcase} 

\begin{figure}
\epsscale{1.}
\includegraphics[width=\columnwidth]{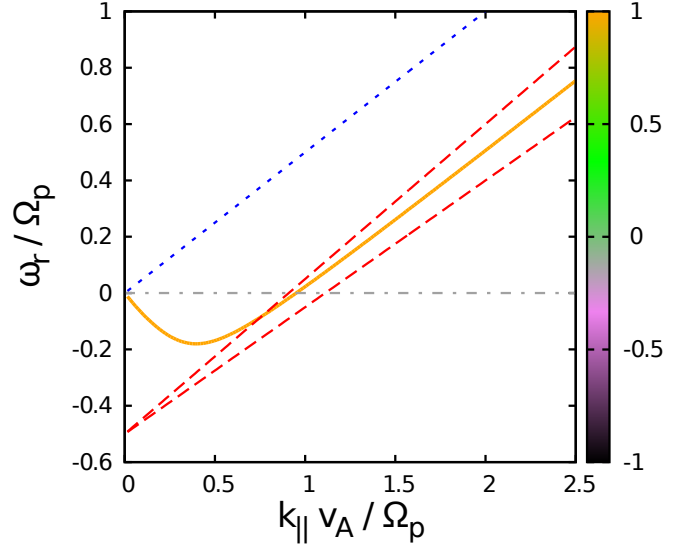}
\caption{Dispersion and polarization for $U_{\mathrm i}=0.5v_{\mathrm
    A}$, $\eta=0.05$, and $\theta=0^{\circ}$. The color coding shows
  the polarization $\mathcal P$. Only branch C$^-$ is shown. The two red lines are the resonance lines for
 $n=-1$. The dotted blue line shows the maximum frequency
  $\omega_{\mathrm r}=k_{\parallel}U_{\mathrm i}$, and the
  triple-dotted gray line is the minimum frequency $\omega_{\mathrm
    r}=0$. \label{fig_first_par}}
\end{figure}

The Landau resonance $n=0$ can be excluded for fast modes and
cyclotron modes when $k_\perp = 0$ because $E_{kz} = 0$ (so that
Landau damping is absent) and because transit-time damping (the other
type of $n=0$ resonance) requires~$k_\perp \neq 0$~\citep{stix92}. The
counter-propagating alpha-cyclotron mode (branch $\mbox{C}^-$ in
Figure~\ref{fig_dispersion}) appears to be a candidate for an
instability driven by the $n=-1$ resonance with the alpha particle
beam, because it satisfies the four criteria for instability listed
above when $\omega_{k\rm r} > 0$, even for modest beam speeds such as
$U_{\rm i} = 0.5 v_{\rm A}$, which can be seen with the aid of
Figure~\ref{fig_first_par}. However, this mode has negative energy
when $\omega_{k\rm r} > 0$, as illustrated in
Figure~\ref{fig_negenerg}, and is therefore stable.

If $\beta_{\parallel \rm i}$ is
sufficiently large, then alpha particles can satisfy the $n=+1$ resonance
condition with the Alfv\'en/proton-cyclotron  wave (branch~A in Figure~\ref{fig_dispersion}) and potentially drive this mode unstable. To the
best of our knowledge, such a parallel Alfv\'en instability has not
been described previously in the literature, although it might play an
important role in the solar wind. We note, however, that our approximations break down when $\beta\gtrsim 1$, because we have approximated the wave frequencies
using the cold-plasma dispersion relation. In order to investigate
parallel Alfv\'en/cyclotron instabilities at large $\beta_{\parallel
  \rm i}$ rigorously, a more sophisticated approach based on the full
hot-plasma dispersion relation would be needed.

\begin{figure*}
\epsscale{1.}
\includegraphics[width=\textwidth]{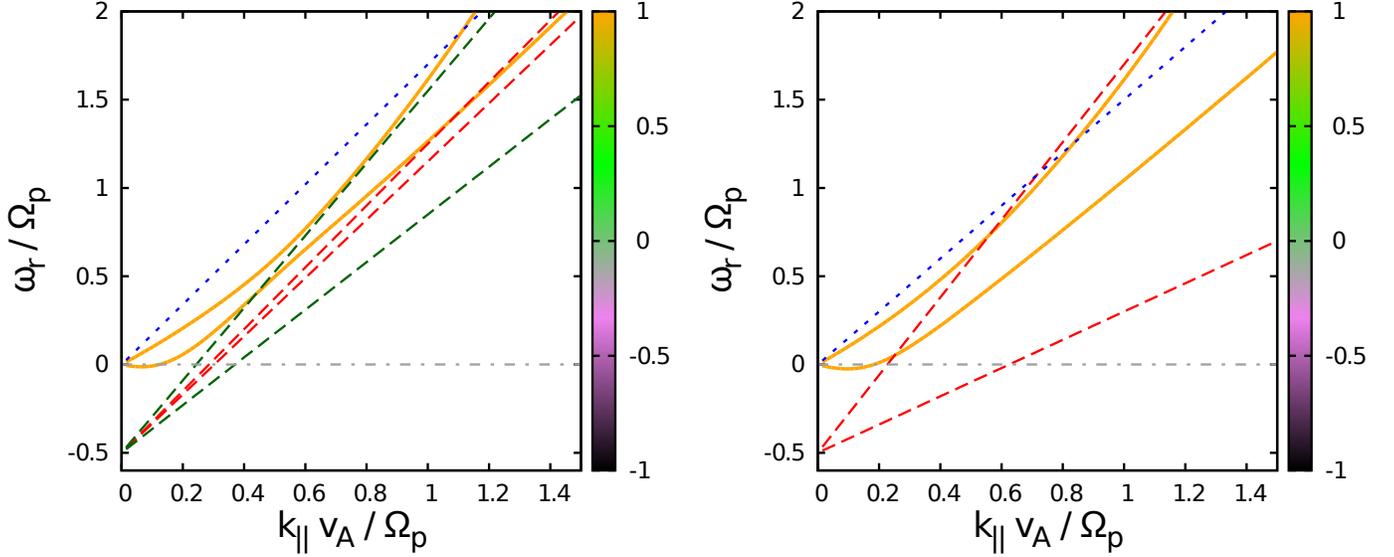}
\caption{Dispersion relation and polarization of the branches B and C$^-$ for $\eta=0.05$ and $\theta=0$. The color coding shows the polarization $\mathcal P$. The resonance lines for $n=-1$ are shown for different thermal speeds. The dotted blue line shows the maximum frequency $\omega_{\mathrm r}=k_{\parallel}U_{\mathrm i}$, and the triple-dotted gray line is the minimum frequency $\omega_{\mathrm r}=0$.  Left: $U_{\mathrm i}=1.7v_{\mathrm A}$. The thermal spread corresponds to $\Delta v_{\parallel}=2v_{\mathrm{thi}}=0.05v_{\mathrm A}$ for the red resonance lines and to $\Delta v_{\parallel}=2v_{\mathrm{thi}}=0.35v_{\mathrm A}$ for the green resonance lines. Right: $U_{\mathrm i}=1.5v_{\mathrm A}$ and $\Delta v_{\parallel}=2v_{\mathrm{thi}}=0.7v_{\mathrm A}$. \label{fig_second_par}}
\end{figure*}

Although the counter-propagating alpha-cyclotron branch C$^-$ is stable when $k_\perp
= 0$, the fast-magnetosonic/whistler mode becomes unstable at sufficiently
large~$U_{\rm i}$ due to the $n=-1$ resonance with the alpha particles.  When
$\Delta v_{\parallel}=2v_{\mathrm{thi}} = 0.35v_{\mathrm A}$, the minimum value of $U_{\rm i}$ for this
instability is $\simeq 1.7 v_{\rm A}$. With the given alpha-particle density, this situation corresponds to a plasma with $\beta_{\parallel \mathrm i}\simeq 0.006$. These parameters are illustrated on the left-hand side in
Figure~\ref{fig_second_par}, and it can be seen from this figure that this
instability satisfies all four criteria in the above list. This instability
corresponds to the parallel magnetosonic instability discussed by \citet{daughton98}
and \citet{gary00b}, and sets in at a typical wavenumber of order the inverse
proton inertial length~$\Omega_{\rm p}/v_{\rm A}$.  The left-hand side in Figure~\ref{fig_second_par} furthermore shows that at this drift speed but at a lower thermal speed the resonance lines do not intersect with the plot of the dispersion relation of the magnetosonic branch. This lower thermal speed represents a situation with a lower $\beta_{\parallel \mathrm i}$ provided the fractional number density $\eta$ is constant.
As $\Delta v_{\parallel}$ is decreased below~$0.3v_{\mathrm A}$, it becomes increasingly
difficult to excite this instability for two reasons. First, the value of
$U_{\rm i}$ needs to be increased simply in order to get the $n=-1$ resonance
line to intersect the dispersion-relation curve for the forward-propagating
fast-magnetosonic/whistler mode. Second, at these larger values of $U_{\rm i}$,
the fast-magnetosonic/whistler branch B merges with the counter-propagating
alpha-cyclotron branch C$^-$ throughout a finite interval of $k_\parallel$ (i.e., they
have the same value of~$\omega_{k\mathrm r}$). Within this interval,
$W<0$ for both modes, and any intersection of an $n=-1$ resonance line with
either mode's dispersion relation is unable to generate an instability according
to the above four criteria.

 As $v_{\mathrm{thi}}$ is increased, our quasilinear/cold-plasma analysis suggests that the threshold
for this instability moves to smaller~$U_{\rm i}$, because larger alpha-particle
thermal speeds enable the alpha-particle $n=-1$ resonance lines to intersect
with the fast-magnetosonic/whistler dispersion relation at smaller values
of~$U_{\rm i}$. This situation is shown on the right-hand side of Figure~\ref{fig_second_par}. With $\eta=0.05$, this situation corresponds to $\beta_{\parallel \mathrm i}\simeq 0.02$. The growth rate dependence of the magnetosonic instability on the thermal speed of the beam has been calculated by \citet{daughton98} for a proton-beam plasma. In qualitative agreement with our arguments, they have shown that the growth rate increases for higher temperatures of the beam.

\subsection{Oblique Case ($k_\perp \neq 0$)}
\label{sec:oblique} 

A moderate obliquity does not significantly change the structure of
the dispersion branches A, B, and $\mbox{C}^\pm$ in
Figure~\ref{fig_dispersion}. It does, however, change the
polarization. When $k_\perp \neq 0$, the waves are in general
elliptically polarized and can interact with particles via both the
$n=+1$ and $n=-1$ cyclotron resonances. Furthermore, Landau-resonant
interactions (with $n=0$) are now possible, because $E_{kz}$ becomes
nonzero for oblique waves, and the nonzero value of $k_\perp$ allows for 
transit-time damping.

The magnetosonic instability that was present in the parallel case
(Section~\ref{sec:parcase}) is also present in the oblique
case. However, whereas the parallel magnetosonic mode resonantly
interacts with the alpha particles only through the $n=-1$ cyclotron
resonance, the oblique magnetosonic mode can interact with alpha
particles through either the $n=-1$ cyclotron resonance or the $n=0$
Landau resonance. For low drift speeds $U_{\mathrm i}\gtrsim v_{\mathrm A}$, the resonance condition with $n=0$ is fulfilled at lower values for $k_{\parallel}$ than the resonance condition with $n=-1$. The two resonance conditions are, therefore, expected to act on the magnetosonic mode in different wavenumber regimes. In the small-wavenumber limit, the fast-wave
dispersion relation is $\omega_{k \rm r} = kv_{\rm A}$, and the
instability condition $U_{\rm i} > \omega_{k \rm r}/k_\parallel$
becomes $U_{\rm i} > v_{\rm A}/\cos\theta$. This argument suggests that the threshold value of the drift speed for the oblique magnetosonic instability is higher at larger $\theta$. 

\begin{figure}
\epsscale{1.}
\includegraphics[width=\columnwidth]{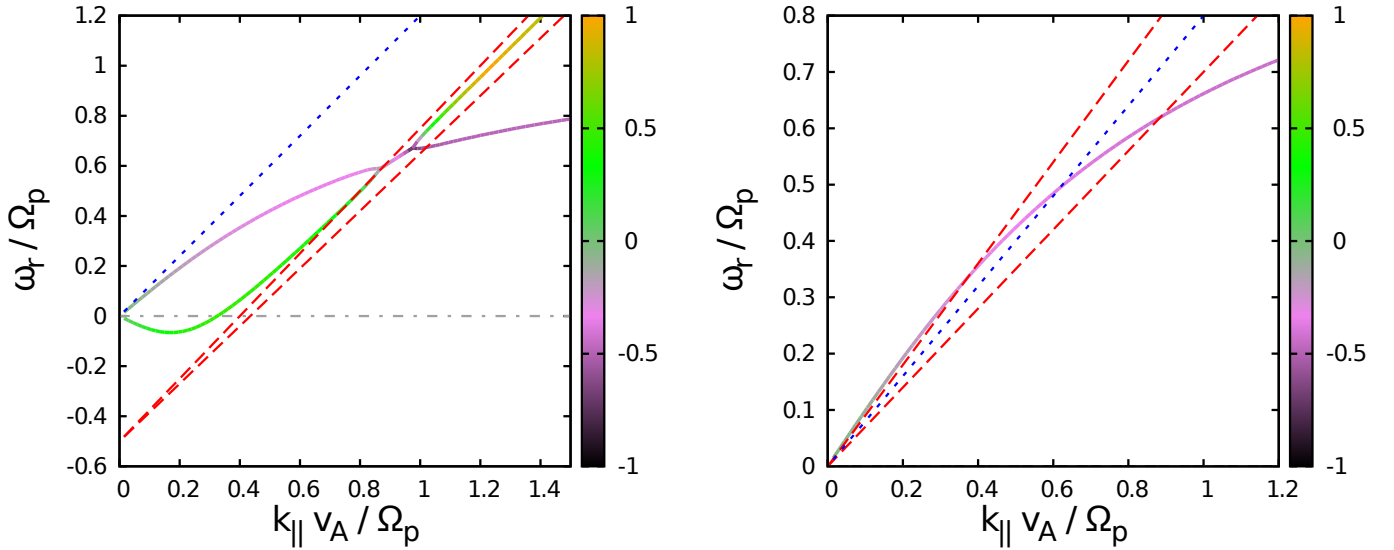}
\caption{Dispersion relation and polarization for the wave branches A and C$^-$. The same parameter set as in Figure~\ref{fig_alpha_res} is used. The color coding shows the value of the polarization function $\mathcal P$. \label{fig_alfven1}}
\end{figure}

The intersection of the $n=-1$ resonance line with the dispersion
relation for proton-cyclotron waves can fulfill constraint 2 for
oblique waves due to their elliptical polarization. To fulfill
constraint 4, the drift speed should be larger than $U_{\mathrm
  i}\approx 1.2v_{\mathrm A}$. Then the intersection of the resonance
line and the dispersion relation for oblique Alfv\'en/proton-cyclotron
waves occurs at low
enough frequencies and wavenumbers ($k_{\parallel} v_{\mathrm
  A}/\Omega_{\mathrm p}\lesssim 1$) that proton damping can be
ignored. This is the situation in Figure~\ref{fig_alfven1}, which also shows
the wave polarization.
We identify this instability with the Alfv\'en I instability discussed
by \citet{gary00}. It begins at $U_{\mathrm i}\gtrsim 1.2v_{\mathrm
  A}$. There is a finite interval of $k_\parallel$ in which the proton
cyclotron wave and the counter-propagating alpha-cyclotron wave
(branches A and $\mbox{C}^-$ in Figure~\ref{fig_dispersion}) merge ---
i.e., in which the two waves have the same value of~$\omega_{k \rm
  r}$. Within this wavenumber interval, both waves have negative
energy, and neither is driven unstable by resonant
particles.\footnote{Within this wavenumber interval, the wave
  frequency is complex, and for one of the two branches the solution
  to the cold-plasma dispersion relation has a positive imaginary
  part. We neglect this type of instability in our discussion for the
  reasons discussed in the Introduction.} 

\begin{figure}
\epsscale{1.}
\includegraphics[width=\columnwidth]{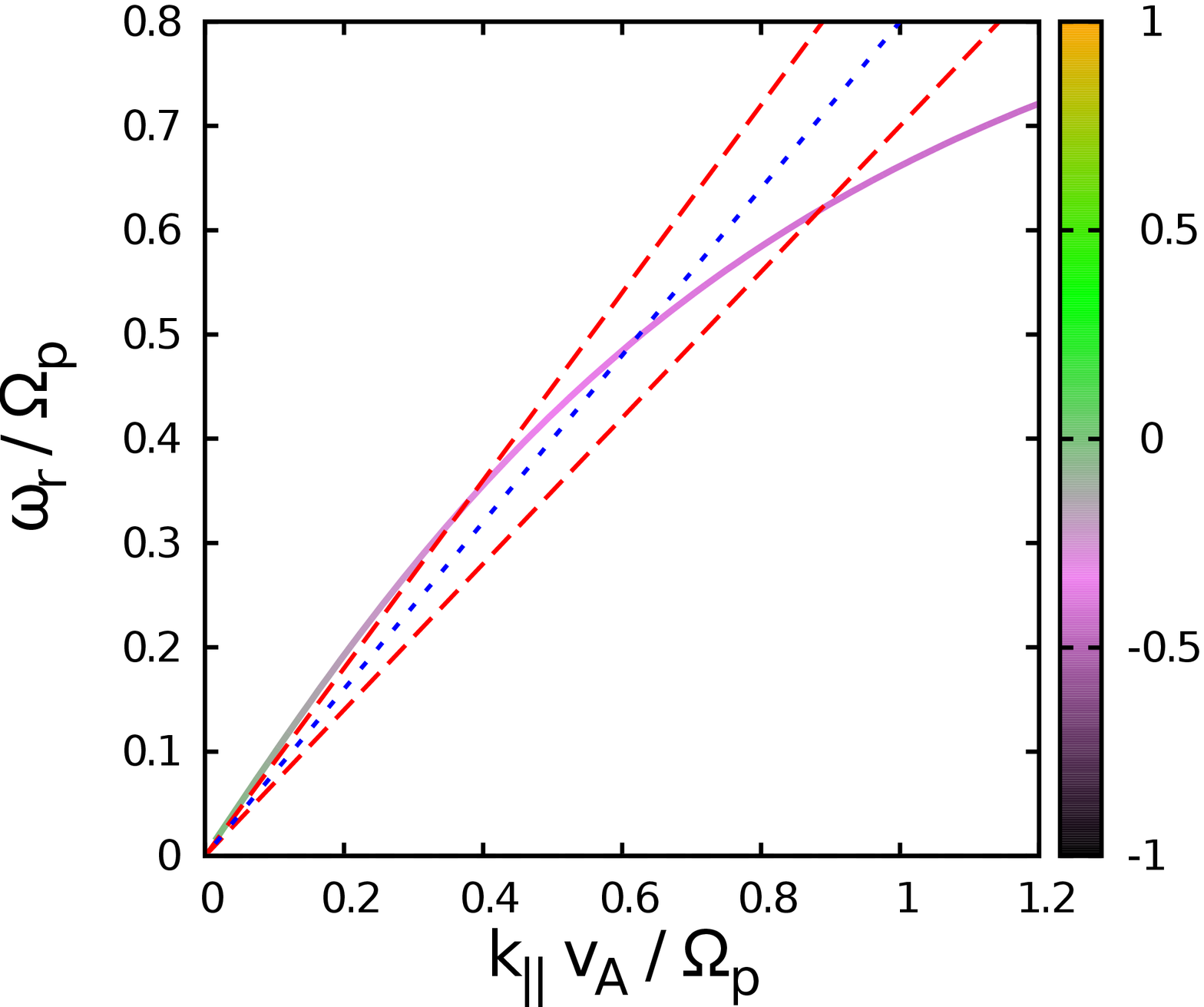}
\caption{Dispersion relation and polarization for the wave branch A with $\theta=45^{\circ}$, $\eta=0.05$, and $U_{\mathrm i}=0.8v_{\mathrm A}$. The color coding shows the value of the polarization function $\mathcal P$. The resonance lines represent the Landau resonance with $n=0$ and $\Delta v_{\parallel}=2v_{\mathrm{thi}}=0.1v_{\mathrm A}$. Only resonance lines below the line $\omega_{\mathrm r}=k_{\parallel}U_{\mathrm i}$ can drive the Alfv\'en II instability according to criterion 3.\label{fig_alfven2}}
\end{figure}

Another instability that becomes possible at oblique propagation is
the so-called Alfv\'en II instability. We identify this instability
with the forward-propagating proton-cyclotron mode, when this mode is
driven unstable by Landau-resonant interactions with the
alpha-particle beam. According to the third of the four criteria laid
out at the beginning of Section~\ref{sec:thresholds}, this instability
requires that $U_{\rm i} > \omega_{k\rm r}/k_\parallel$. The parallel
phase velocity $\omega_{k \rm r}/k_\parallel$ of the proton-cyclotron
wave is $v_{\rm A}$ for $k_\parallel \ll \Omega_{\rm p}/v_{\rm A}$ and
decreases towards zero at $k_\parallel \rightarrow \infty$. Thus, the
condition $U_{\rm i} > \omega_{k \rm r}/k_\parallel$ is always
satisfied for positive $U_{\rm i}$ provided that $k_\parallel$ is
sufficiently large. However, if $k_\parallel$ is too large, criterion
4 in the above list is not satisfied, because $\omega_{k\rm r} \sim
\Omega_{\rm p}$ and the wave is damped by cyclotron resonant
interactions with protons. In order to avoid proton cyclotron damping,
$k_\parallel$ must be $\lesssim \Omega_{\rm p}/v_{\rm A}$ (the precise
upper limit depends upon the value of $\beta_{\parallel \rm p}$ for
the protons). In order to have $U_{\rm i} > \omega_{k \rm
  r}/k_\parallel$ at $k_\parallel < \Omega_{\rm p}/v_{\rm A}$, $U_{\rm
  i}$ must exceed~$\simeq 0.8 v_{\rm A}$, which is the approximate
threshold-value of~$U_{\rm i}$ for this instability.  This case is shown in Figure~\ref{fig_alfven2}.
Criterion three in
the above list of four criteria, combined with the Landau resonance
condition $\omega_{k \rm r}/k_\parallel = v_\parallel$, imply that
only alpha particles with $v_\parallel < U_{\rm i}$ can drive the
Alfv\'en~II instability. As $U_{\rm i}$ is increased beyond~$v_{\rm
  A}$, fewer and fewer alpha particles can undergo a Landau resonance
with the proton-cyclotron wave, and thus there is an upper limit
on~$U_{\rm i}$ for the Alfv\'en~II instability.  As $\beta_{\parallel
  \rm i}$ increases for the alpha particles, this upper limit
increases.

There is evidence in previous treatments of beam instabilities that another kinetic effect may weaken or suppress the oblique instabilities in certain parameter regimes. It was found by \citet{montgomery76} in a proton-beam plasma that the oblique Alfv\'enic instability (which corresponds to Alfv\'en II according to \citealt{daughton98}) shows a higher growth rate at greater values of $T_{\mathrm e}/T_{\mathrm p}$, whereas the cyclotron-resonant magnetosonic instability does not. \citet{montgomery76} interpreted this result as a consequence of Landau-damping of these modes by electrons. At higher electron temperatures, the gradients of the electron distribution function in velocity space become smaller since the width of the distribution function increases. The influence of electron Landau damping is determined by these gradients and, therefore, mitigated at higher electron temperatures. In our model, the Landau-resonance condition for electrons would correspond to resonance lines that are very widely spread due to their large thermal speed. Therefore, almost all waves with $\mathcal P_z\neq 0$ in our frequency range can undergo electron Landau damping. The Alfv\'en I instability may not be as strongly influenced by this effect as the Landau-resonant instabilities since the Alfv\'en I instability generally shows higher growth rates in the numerical solutions of the hot-plasma dispersion relation under typical parameters \citep[e.g.,][]{gary00}. This effect may also be the reason why \citet{montgomery76} find that the oblique magnetosonic instability has a significant growth rate over only a very small range of the plasma parameters, and why \citet{gary00} do not discuss this mode for a proton-alpha plasma. The details of this effect, however, require a deeper study and are beyond the scope of this work.

\section{Discussion and Conclusions}
\label{sec:discussion}

In this paper, we use quasilinear theory to explain the thresholds for
Alfv\'en/ion-cyclotron and magnetosonic/whistler instabilities driven
by resonant interactions with an alpha-particle beam in low-$\beta$ plasmas. Our arguments
rely upon three properties of the wave modes in question: the real
part of the wave frequency~$\omega_{k \rm r}$, the polarization of the
wave electric field, and the sign of the wave energy density~$W$. (See
Section~\ref{sec:W} for a discussion of negative-energy waves.) To
obtain approximate values for these quantities as functions of the
wavevector~$\vec k$, we have used the cold-plasma approximation,
taking into account the finite electron mass. 

The general considerations that determine whether a wave is driven
unstable by resonant interactions with a beam of alpha particles are
the following. First, in order for a wave to be unstable at some value
of~$k_\parallel$, it can not undergo cyclotron resonant interactions
with the thermal protons at that~$k_\parallel$, since such
interactions would strongly damp the wave.  Second, the wave and alpha
particles must satisfy the wave--particle resonance condition,
Equation~(\ref{resecond}). Third, the resonant wave--particle
interactions must cause the alpha particles to lose energy, so that
the wave can gain energy and grow. (This statement assumes that $W>0$
--- more details on the stability and instability of negative-energy waves are given in
Section~\ref{sec:thresholds}). As we show in
Section~\ref{sec:thresholds} and in the Appendix, resonant wave--particle interactions
cause alpha particles with a distribution function that is isotropic about their mean velocity to lose energy when~$0 < \omega_{k\mathrm r}/k_\parallel
< U_{\rm i}$, where $U_{\rm i}$ is the speed of the alpha particle
beam, and all quantities (e.g., $\omega_{k\mathrm r}$, $U_{\rm i}$) are
measured in the proton rest frame. Fourth, the wave polarization must
satisfy the following conditions. In order for a wave to be driven
unstable by an $n=+1$ ($n=-1$) cyclotron resonance, $E_{k,\rm l}$
($E_{k, \rm r}$) must be nonzero. When $k_\perp = 0$, a wave can be
driven unstable by the $n=0$ Landau resonance only when~$E_{kz}
\neq 0$.

When we apply these considerations to parallel waves with $k_\perp =
0$, we find that alpha particles are unable to drive resonant
instabilities when $\Delta v_{\parallel}=2v_{\mathrm{thi}}$ is much smaller
than~$0.35$. When $\Delta v_{\parallel} \simeq 0.35$, our criteria
suggest that the magnetosonic/whistler wave becomes unstable via the
$n=-1$ resonance when~$U_{\rm i}$ exceeds $\sim 1.7 v_{\rm A}$,
consistent with previous numerical results \citep{li00,gary00b}. Our arguments
suggest that this minimum threshold on $U_{\rm i}$ decreases slightly
as the parallel thermal speed of the ions increases further. Our arguments also
suggest that as $v_{\mathrm{thi}}$ (and therefore $\beta_{\parallel \mathrm i}$) increases, the alpha particles eventually drive the
proton-cyclotron wave (branch A in Figure~\ref{fig_dispersion})
 unstable in the~$n=+1$ resonance.   We note,
however, that our conclusions about the high-$\beta_{\parallel \rm i}$
case must be treated with caution, because of our use of the
cold-plasma dispersion relation.

When we apply the above instability criteria to oblique waves with
$k_\perp \neq 0$, we find that the magnetosonic mode can still be unstable, but that it can now be rooted in either the cyclotron resonance ($n=-1$)
or the Landau resonance ($n=0$).  The minimum threshold to excite the
oblique magnetosonic instability via the Landau resonance is somewhat
smaller than for the parallel magnetosonic instability, and also less
dependent on~$v_{\mathrm{thi}}$. The wavenumbers at which the mode is unstable when driven by the $n=0$ resonance are smaller than the unstable wavenumbers for oblique magnetosonic instabilities driven by the $n=-1$ resonance. At these smaller wavenumbers, the dispersion relation of the magnetosonic mode for the $n=0$ resonant instability can be approximated by $\omega_{k\mathrm r}\approx k_{\parallel}v_{\mathrm A}/\cos\theta$ which leads to an instability threshold of $U_{\mathrm i} \gtrsim v_{\mathrm A}/\cos \theta$ according to the instability criterion 3 introduced in Section~\ref{sec:thresholds}.  At oblique propagation, there
are also two types of Alfv\'en/cyclotron instabilities that are not
present at parallel propagation. The first of these, the ``Alfv\'en
I'' instability, is driven by the $n=-1$ cyclotron resonance. Our
analysis suggests that this instability requires $U_{\rm i} \gtrsim
1.2 v_{\rm A}$, which is consistent with the numerical results of
\cite{gary00} for a plasma with $T_{\mathrm e}/T_{\mathrm p}=4$. Our physical interpretation of this instability
threshold is that it is the minimum value of~$U_{\rm i}$ for which
alpha particles can resonate with the proton-cyclotron wave (branch~A
in Figure~\ref{fig_dispersion}) through the~$n=-1$ resonance at a
wavenumber~$k_\parallel$ that is sufficiently small to avoid strong
cyclotron damping by the protons.  The second oblique instability, the
so-called Alfv\'en~II instability \citep{daughton98,gary00}, is driven
by the $n=0$ Landau resonance. Our arguments suggest that this
instability requires that $U_{\rm i} \gtrsim 0.8 v_{\rm A}$,
consistent with the numerical results of \cite{gary00} for the case in
which $T_{\rm e}/T_{\rm p} = 4$.  This lower threshold on~$U_{\rm i}$
is the minimum beam speed for which the proton-cyclotron wave
(branch~A in Figure~\ref{fig_dispersion}) can undergo a Landau ($n=0$)
resonance with the alpha particles at a~$k_\parallel$ that is
sufficiently small to avoid strong proton cyclotron damping.  We note, however, as discussed in
  Section~\ref{sec:oblique}, that the Alfv\'enic instabilities and the oblique magnetosonic instability may weaken
  as $T_{\rm e}/T_{\rm p}$ decreases because of the more efficient electron Landau damping of these modes. This effect may lead to a weakening or suppression of these instabilities in the
  fast solar wind, where $T_{\rm e}$ is typically smaller than~$T_{\rm
    p}$ by a factor of a few.  A summary of our findings for the
instability thresholds of oblique and parallel waves is given in
Table~\ref{tab_summary}.  

\begin{deluxetable}{lcc}
\tablecaption{Summary of the Unstable Modes and Their Thresholds\label{tab_summary}}
\tablehead{\colhead{Name} & \colhead{Resonance} & \colhead{Threshold}}
\startdata 
Parallel magnetosonic & $n=-1$ & $\sim 1.7v_{\mathrm A}$  \\
Oblique magnetosonic\footnote{The oblique magnetosonic mode can be driven unstable by two different resonances. The $n=0$ resonance has the lower threshold given in this table and occurs then at lower wavenumbers compared to the $n=-1$ resonance.}& $n=0$ or $n=-1$ & $\gtrsim v_{\mathrm A}/\cos\theta$  \\
Alfv\'en I & $n=-1$ & $\sim 1.2 v_{\mathrm A} $  \\
Alfv\'en II & $n=0$ & $\sim 0.8 v_{\mathrm A} $ 
\enddata
\tablecomments{ The three oblique instabilities may depend on $T_{\mathrm e}/T_{\mathrm p}$ as described in Section~\ref{sec:oblique}. }
\end{deluxetable}

The principal utility of our analysis is that it provides a
conceptual framework for understanding previous numerical results on
the different instability thresholds, including the  finding
that the instability threshold drops from~$U_{\rm i} \gtrsim 1.7 v_{\rm
  A}$ for parallel propagation to~$U_{\rm i}\gtrsim 0.8 v_{\rm A}$ for
oblique propagation. Our discussion also describes the different
conditions under which cyclotron resonances~(either $n=+1$ or $n=-1$)
and the Landau resonance~$n=0$ are important.

The wider importance of this topic is that beam-driven instabilities
 place an interplanetary `speed limit' on alpha-particle beams. As
demonstrated in numerous previous studies \citep{marsch87,daughton99,gary00,goldstein00,reisenfeld01,kaghashvili03,gary03,hellinger03,hellinger11}, when $U_{\rm i}$ exceeds the threshold for a beam instability, exponential growth of the unstable
wave and the resulting wave--particle interactions decelerate the
beam and limit the beam speed to approximately the marginally stable
value. If this scenario is correct, then it is the instability with
the smallest $U_{\rm i}$-threshold that is most important for limiting
the beam speed in the solar wind. At low~$\beta$, the instability with
the lowest threshold is the Alfv\'en~II instability, at least
when~$T_{\rm e}/T_{\rm p}$ is sufficiently large (see discussion in
Section~\ref{sec:oblique}). At $T_{\rm e}/T_{\rm p} \lesssim 1$, the
Alfv\'en~II instability may be more difficult to excite, in which case
the instability with the lowest instability threshold could become the
Alfv\'en~I instability.  Regardless of which instability
has the lowest beam-speed threshold, each of the instabilities that we
have investigated has a threshold-value of~$U_{\rm i}$ that is
normalized to the Alfv\'en speed. Because $v_{\rm A}$ decreases with
increasing heliocentric distance~$r$ outside of the solar corona, if
instabilities limit the value of~$U_{\rm i}$ to $\simeq v_{\rm A}$ in
the solar wind, then they lead to ongoing deceleration of the
alpha-particle beam at $r>r_{\rm min}$, where $r_{\rm min}$ is the
radius at which $U_{\rm i}$ first reaches a value~$\sim v_{\rm A}$ as
the alpha particles flow away from the Sun. This scenario of
deceleration by beam-driven instabilities is consistent with {\em
  Helios} measurements showing that $U_{\rm i} \lesssim v_{\rm A}$ at
$0.3 \mbox{ AU} < r < \mbox{ 1 AU}$~and \emph{Ulysses} observations of proton beams at up to $ r = \mbox{ 3 AU}$ \citep{marsch82,goldstein00}.  

Another reason that beam-driven instabilities could be important in the solar wind is that they could lead to an inverse cascade of wave power from
large~$k_\parallel$ to smaller~$k_\parallel$. In this picture, the ion
beams generate Alfv\'en/cyclotron waves at $\omega \sim \Omega_{\rm
  p}$ and $k_\parallel \sim \Omega_{\rm p}/v_{\rm A}$. In the MHD
limit ($k \ll \Omega_{\rm p}/v_{\rm A}$), an Alfv\'en wave can decay
into another Alfv\'en wave, which propagates in the opposite
direction, and a slow mode wave, which propagates in the same
direction as the mother wave, a process called ``the parametric
instability'' \citep{sagdeev69,cohen74}.  The daughter Alfv\'en wave
has a slightly smaller $k_\parallel$ and~$\omega$ than the mother
wave, and over time the parametric instability leads to an inverse
cascade, as described by \citet{cohen74}, \citet{kuznetsov01}, and
\citet{chandran08} for a weak turbulence scenario at low beta and
$k\ll \Omega_{\rm p}/v_{\rm A}$. Hybrid numerical
simulations have shown that a similar parametric instability and
inverse cascade occur at $k\sim \Omega_{\rm p}/v_{\rm A}$ \citep{markovskii09,matteini10,verscharen12b}. An inverse cascade at $k \sim
\Omega_{\rm p}/v_{\rm A}$ feeds energy into the low-frequency MHD
regime, where the MHD inverse cascade can continue transferring energy
to lower frequencies. This scenario is consistent with observations
from the \emph{ACE} satellite \citep{stawarz10}, which show evidence for
energy transfer from small to large scales when the solar wind is in a
state of high cross-helicity.  An inverse cascade could produce
``slab-like'' magnetic fluctuations in the solar wind with
$k_\parallel \lesssim k_\perp$ over a broad range of parallel
wavenumbers, which could lead to efficient scattering of energetic
particles~\citep{jokipii66,kulsrud69}. In contrast, the quasi two-dimensional
fluctuations with $k_\perp \gg k_\parallel$ produced by the forward
Alfv\'en-wave energy
cascade~\citep{shebalin83,goldreich95,oughton95,verscharen12} are
highly ineffective at scattering energetic
particles~\citep{bieber94,chandran00b,yan02}. Further investigations
of this inverse cascade will be an interesting task for a future
study, which could eventually connect plasma instabilities and
turbulence to the transport of energetic particles in the solar wind.

\acknowledgments

We appreciate helpful discussions with Joe Hollweg, Marty Lee, and
Phil Isenberg.  This work was supported in part by grant NNX11AJ37G
from NASA's Heliophysics Theory Program, NASA grant NNX12AB27G,
NSF/DOE grant AGS-1003451, and DOE grant DE-FG02-07-ER46372.

\appendix 
\section{Proof that an Isotropic Distribution of Beam Particles Loses Energy From Resonant Interactions with Waves satisfying \lowercase{$0<\omega_{k\mathrm r}/k_{\parallel}<\uppercase{U}_{\mathrm i}$}}

In this Appendix, we consider the way that the particle kinetic energy
changes in response to resonant wave-particle interactions. The
kinetic energy density in the proton rest frame of particle species~$s$ is ${\mathcal E} = 0.5
m_s \int f_s \vec v^2 d^3 v$. Since we have taken $f_s$ to be independent
of spatial position~$\vec x$, $\mathcal E$ is a function of~$t$ alone,
and we can write
\begin{equation}
\frac{\mathrm d \mathcal E}{\mathrm d t}=\frac{1}{2}m_s\int \vec v ^2\frac{\partial f_s}{\partial t}\mathrm d^3 v.
\label{eq:dEdt0} 
\end{equation}
Using Equation~(\ref{qldiff}) to describe the quasilinear evolution of~$f_s$,
we can rewrite Equation~(\ref{eq:dEdt0})  as
\begin{multline}\label{dedt}
\frac{\mathrm d \mathcal E}{\mathrm d t}=\lim _{V\to \infty}\sum \limits_{n=-\infty}^{+\infty}\frac{q_s^2}{8 \pi m_s} \int\mathrm d^3k \int\limits_{-\infty}^{+\infty} \mathrm dv_{\parallel} \\
\times \int\limits_0^{+\infty}  \mathrm dv_{\perp} \frac{\left(v_{\perp}^2+v_{\parallel}^2\right)}{V}Gv_{\perp} \\
\times \delta(\omega_{k\mathrm r}-k_{\parallel}v_{\parallel}-n\Omega_s)\left|\psi_{n,k} \right|^2Gf_s.
\end{multline}
We assume that $f_s$ is isotropic about the beam velocity $U_{\mathrm i}\hat{\vec e}_z$ and monotonically decreasing with distance from $U_{\mathrm i}\hat{\vec e}_z$ in velocity space, i.e.
\begin{align}
f_s&=f_s\left((\vec v-U_{\mathrm i}\hat{\vec e}_z)^2\right)\label{fsiso},\\
f_s^{\prime}&\le0,
\label{eq:fsleq0} 
\end{align}
where the prime indicates the derivative with respect to the argument of $f_s$ in Equation~(\ref{fsiso}). 
Applying the $G$-operator to $f_s$ yields
\begin{equation}
Gf_s=2v_{\perp}\left(1-\frac{k_{\parallel} U_{\mathrm i}}{\omega_{k\mathrm r}}\right)f_s^{\prime}.
\end{equation}
The first $G$-operator in Equation~(\ref{dedt}) can be eliminated by integration by parts, leading to
\begin{equation}\label{ikdk}
\frac{\mathrm d \mathcal E}{\mathrm d t}=\int   I_k \; \mathrm d^3k,
\end{equation}
where
\begin{multline}\label{ik}
I_k= \left(1-\frac{k_{\parallel} U_{\mathrm i}}{\omega_{k\mathrm r}}\right) \left[
\lim _{V\to \infty} \:\frac{q_s^2}{2 \pi m_s} \,\sum \limits_{n=-\infty}^{+\infty} \;\;\,\int\limits_{-\infty}^{+\infty} \mathrm dv_{\parallel} \right. \\
\times \left.\int\limits_0^{+\infty} \mathrm dv_{\perp}   \frac{v_{\perp}^3}{V}\delta(\omega_{k\mathrm r}-k_{\parallel}v_{\parallel}-n\Omega_s)\left|\psi_{n,k} \right|^2 |f_s^{\prime}|\right].
\end{multline}
The quantity $I_k$ describes the contribution to $\mathrm d\mathcal E/\mathrm dt$ from waves at wave vector~$\vec k$.

All terms on the right-hand side of Equation~(\ref{ik}) are positive
semi-definite except for the quantity
\begin{equation}
g=1-\frac{k_{\parallel}U_{\mathrm i}}{\omega_{k\mathrm r}},
\end{equation} 
which can be positive or negative depending on the ratio of
$U_{\mathrm i}$ to $v_{\mathrm{ph}}=\omega_{k\mathrm r}/k_{\parallel}$
and the sign of $v_{\mathrm{ph}}$. The sign of $I_k$ is then negative
if and only if: (1) $g<0$; and (2) $|\psi_{n,k}|^2\neq 0$ and
$f_s^{\prime}\neq 0$ (meaning $f_s^{\prime}< 0$ given
Equation~(\ref{eq:fsleq0})) for at least some subset of the particles that
satisfy the resonance condition. The case in which $f_s^\prime = 0$
for all resonant particles is not relevant to ion beams in the solar
wind, and so we do not consider this case further. The quantity
$|\psi_{n,k}|^2$ can vanish if the wave amplitude vanishes or if
the wave polarization and propagation direction satisfy certain constraints:
e.g., if $n=1$, $k_\perp = 0$, and $E_{k,\mathrm l}=0$. We have
addressed the constraints on wave polarization elsewhere in this paper
(see, e.g., instability criterion~2 in Section~\ref{sec:thresholds}),
and we assume that there are waves that can resonate with the
particles. In the remainder of this Appendix, we thus neglect the case
in which $|\psi_{n,k}|^2$ is identically zero in the integrand within
the brackets of Equation~(\ref{ik}). The sign of $I_k$ is then equal
to the sign of~$g$.  Therefore, if we wish to determine whether waves
at some particular wave vector~$\vec k_1$ act to increase or decrease
the particle kinetic energy, we need only determine the sign of~$g$ at
$\vec k = \vec k_1$, which depends only upon the value of $\omega_{k
  \mathrm r}/k_\parallel$ at $\vec k = \vec k_1$.  The condition $g<0$
is equivalent to our criterion 3 in Section~\ref{sec:thresholds}, namely
$0<\omega_{k\mathrm r}/k_{\parallel}<U_{\mathrm i}$.  We note that if
a spectrum of waves is excited over a range of~$\vec k$ values, then
some waves may act to increase~$\mathcal E$ while others act to
decrease~$\mathcal E$. The sign of $\mathrm d{\mathcal E}/\mathrm dt$ must then be
determined by evaluating the integral in Equation~(\ref{ikdk}).  \\

\bibliographystyle{apj}
\bibliography{oblique_aic_alphas}

\end{document}